\documentclass[twocolumn,aps,prl,showpacs,groupedaddress]{revtex4}

\usepackage{amsmath}
\usepackage{dcolumn}
\usepackage{epsfig}
\usepackage{graphicx}
\usepackage{latexsym}

\usepackage{epstopdf}

\begin{document}

\title{Spin gradient demagnetization cooling of ultracold atoms} 

\author{Patrick Medley}
\altaffiliation{PM and DMW contributed equally to this work.  Present address of PM: Department of Physics, Stanford University, Stanford CA 94305}
\author{David M. Weld}
\email{dweld@mit.edu}
\author{Hirokazu Miyake}
\author{David E. Pritchard}
\author{Wolfgang Ketterle}
\affiliation{MIT-Harvard Center for Ultracold Atoms, Research Laboratory of Electronics, and Department of Physics, Massachusetts Institute of Technology, Cambridge MA 02139}

\begin{abstract}
We demonstrate a new cooling method in which a time-varying magnetic field gradient is applied to an ultracold spin mixture.  This enables preparation of isolated spin distributions at positive and negative effective spin temperatures of $\pm$50 picokelvin. The spin system can also be used to cool other degrees of freedom, and we have used this coupling to cool an apparently equilibrated Mott insulator of rubidium atoms to 350 picokelvin. These are the lowest temperatures ever measured in any system. The entropy of the spin mixture is in the regime where magnetic ordering is expected.
  \end{abstract}


\maketitle 

Attainment of lower temperatures has often enabled discovery of new phenomena, from superconductivity to Bose-Einstein condensation.  Currently, there is much interest in the possibility of observing correlated magnetic quantum phases in lattice-trapped ultracold atoms~\cite{lewenstein-review,zwerger-review,duandemlerlukin}.  The relevant critical temperatures are on the order of 200 picokelvin, lower than any previously achieved temperature~\cite{criticalentropy,ho-mott}.  Realization of this temperature scale requires the development of new methods of refrigeration which can be applied to ultracold atoms.  Many such techniques have been proposed~\cite{pomeranchukcoolingfermions,cirac-cooling,umass-bosehubbard,daley-latticecooling,coolingbyshaping,ho-coolingbysqueezing,ho-coolingarxiv,entropyexchange_KRb} but await experimental realization.  The cooling method we demonstrate here opens up a previously inaccessible temperature regime and provides a realistic path to the observation of magnetic quantum phase transitions in optical lattices.

The new cooling method is applied to an optically trapped cloud of cold atoms in a mixture of two internal states with different magnetic moments~\cite{EPAPSnote}.  Application of a strong magnetic field gradient results in almost complete spatial segregation of the two spin components.  The ``mixed region'' of spins between the pure-spin domains has a width which is proportional to the temperature $T$ and inversely proportional to the applied gradient.  Reducing the gradient mixes the two components, and due to the mixing entropy the temperature is dramatically reduced.  Since Mott insulators can be prepared with an entropy per particle much lower than $k_B\ln2$, our scheme is a practical way of creating a low entropy mixture in the regime where magnetic ordering is expected.  This cooling scheme introduces several new concepts.  It implements demagnetization cooling with spin transport instead of spin flips, allows decoupling of the spin and kinetic temperatures, and enables the realization of negative spin temperatures. Long tunneling times in the lattice allow two different implementations of our cooling scheme.

When the gradient is changed faster than the spin relaxation rate, the spin system is effectively isolated from all other degrees of freedom, and very low spin temperatures can be achieved.  Contrastingly, when the gradient is changed slowly enough, the spin system is fully equilibrated and can absorb entropy from other degrees of freedom, cooling the whole sample.  Reduction of the gradient after (before) the optical lattice has been raised realizes the regime of isolated (equilibrated) spins.  

First, we discuss isolated spins, of which atoms in a Mott insulating state are an almost ideal realization. Spin distributions relax by two atoms exchanging sites through a second order tunneling process. The time scale of this relaxation is typically one second, and the gradient can easily be varied much faster. The equilibrium spin distribution depends only on the ratio of the applied gradient $\nabla |\textbf{B}|$ and temperature $T$.  When $\nabla |\textbf{B}|$ is changed the effective temperature of the decoupled spin degrees of freedom (spin temperature) is rescaled proportionally.   This enables the realization of spin distributions with a very low positive (or, if the sign of the gradient is changed, negative) effective temperature.  Negative temperatures can only occur for systems with an upper bound on the energy, and correspond to an inverted Boltzmann distribution with the largest population in the highest-energy state~\cite{reif105}.

\begin{figure}
\begin{center}
\includegraphics[width=0.5\columnwidth]{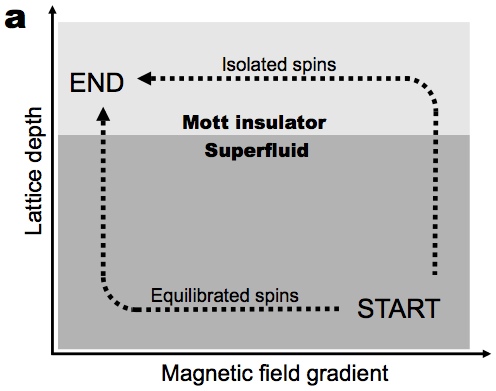}
\includegraphics[width=0.4\columnwidth]{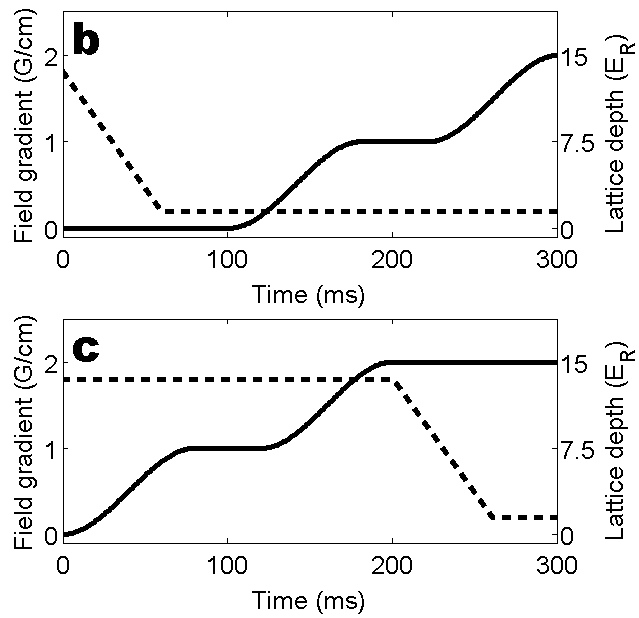}
\caption{Two different cooling protocols realizing the cases of isolated and equilibrated spin systems. \textbf{a:} Experimental ``phase diagram'' of lattice depth vs. applied gradient.  Dashed lines show two different paths which connect the high-gradient superfluid state and the low-gradient Mott insulating state.  \textbf{b:} and \textbf{c:} show the lattice depth (solid line) and gradient strength (dashed line) versus time for the two cases (equilibrated spins and isolated spins, respectively) in panel \textbf{a}. The shape of the lattice rampup is designed to ensure maximum equilibration.
\label{phasediagram}}
\end{center}
\end{figure}
\begin{figure}[tbh]
\begin{center}
\includegraphics[width=0.7\columnwidth]{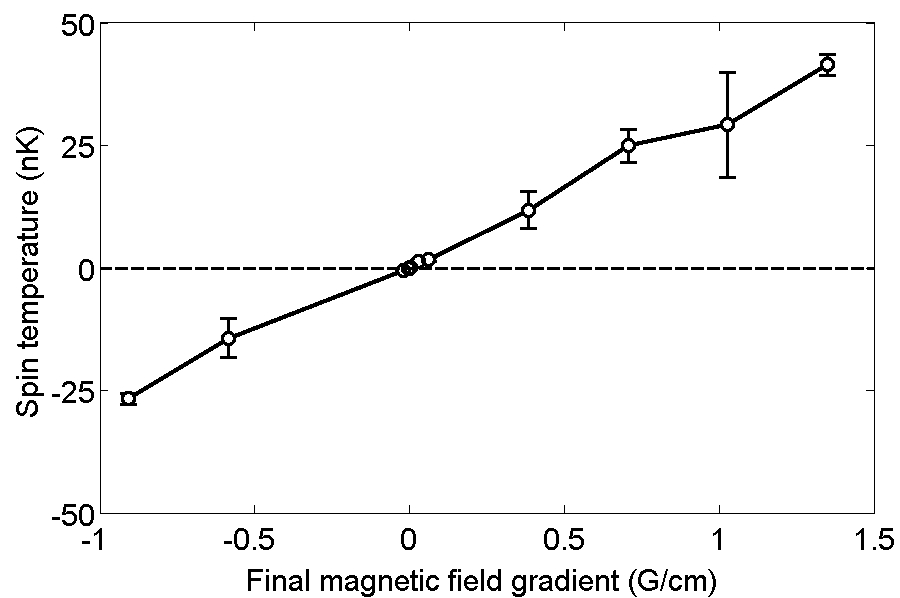}
\caption{Preparation of low positive and negative spin temperatures.  Measured spin temperature versus final gradient, for the case of isolated spins.  Error bars are statistical.
\label{spintemps}}
\end{center}
\end{figure}
\begin{figure}[tb]
\begin{center}
\includegraphics[width=0.9\columnwidth]{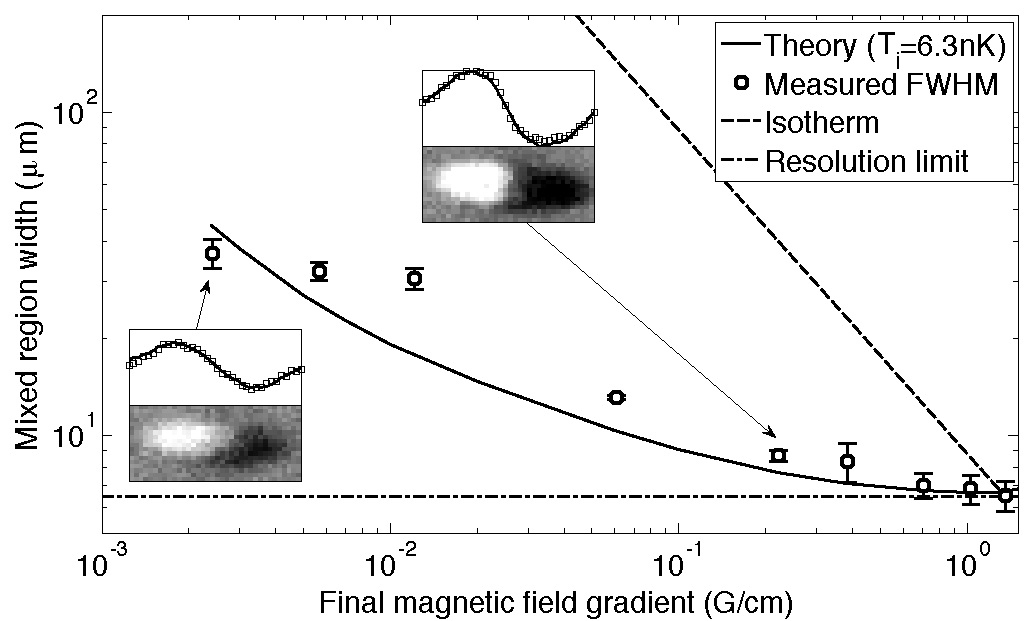}
\caption{Entropy transfer from other degrees of freedom to the spin system.  Circles represent measured width of the mixed region vs. final magnetic field gradient, for the ``equilibrated spins'' protocol (see Fig.~\ref{phasediagram}).  Error bars are statistical.  The dashed line represents the expected behaviour assuming no cooling.  The dash-dotted line shows the minimum resolvable width.  The solid curve is the theoretical prediction, assuming an initial temperature of 6.3 nanokelvin and including the effects of optical resolution (see Ref.~\cite{demagtheory} for details).  Insets show spin images at the indicated points, the corresponding vertically integrated spin profiles (squares), and the fit to the expected form of a $\tanh$ function times the overall density distribution (solid line; see Ref.~\cite{thermometrypaper} for details).  Axis units and grayscales in the two insets are abitrary but identical.  \label{widthvsgrad}}
\end{center}
\end{figure}
\begin{figure}[tb]
\begin{center}
\includegraphics[width=0.9\columnwidth]{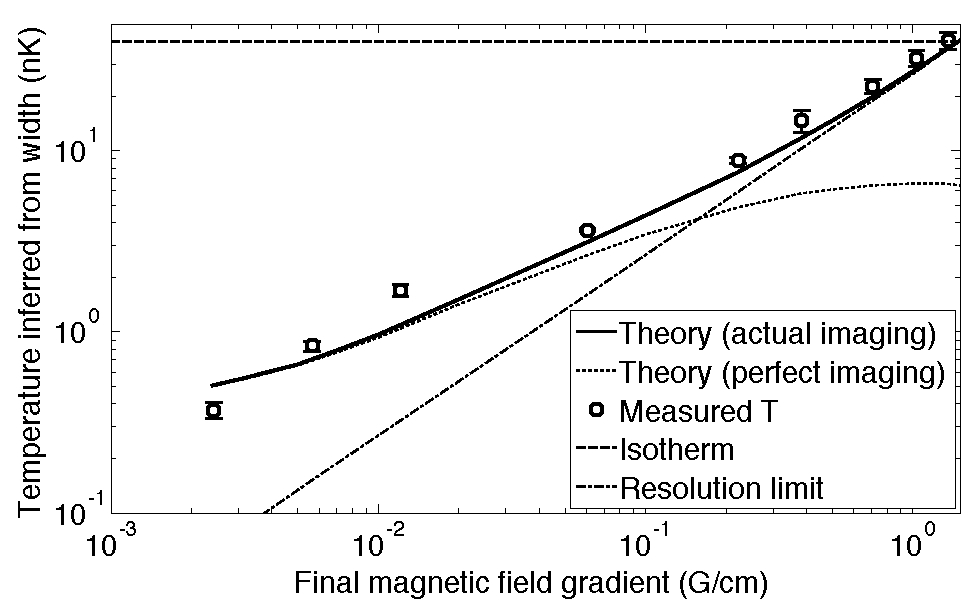}
\caption{Spin gradient demagnetization cooling.  Circles represent measured temperature vs. final magnetic field gradient, for the ``equilibrated spins'' protocol (see Fig.~\ref{phasediagram}).  Error bars are statistical.  These measurements are the same as those shown in Fig.~\ref{widthvsgrad}.  The dashed line follows the isothermal trajectory and the dash-dotted line shows the resolution limit. The solid line is the theoretical prediction, assuming an initial temperature of 6.3 nanokelvin and including the effects of optical resolution.  The dotted line is the theoretical prediction \emph{without} the effects of optical resolution.  \label{Tvsgrad}}
\end{center}
\end{figure}

Figure~\ref{spintemps} shows the results of experiments on \emph{isolated} spins.  Fits to data on equilibrated spins indicate an initial temperature of 6.3 nanokelvin (see Figs.~\ref{widthvsgrad}~and~\ref{Tvsgrad}).  While reduction of the gradient by a factor of 1000 would be expected to reduce the effective temperature of the isolated spins to 6.3 picokelvin, our finite optical resolution only allows us to assert an upper bound of 50$\pm$20 picokelvin.  We have held the spins for up to several seconds in the lattice.  No heating is observed during a one-second hold, but after three seconds the 50 picokelvin distribution is observed to heat to about 70 picoKelvin.  Similarly, at small negative gradients, we observe a negative-temperature distribution with a temperature closer to zero than -50$\pm$20 picokelvin.  Since the total energy is monotonic in $-1/T$, these are among the most extreme thermodynamic states ever measured in any system~\cite{helsinki-spintemps,helsinki-spintemps2}.  Very low condensate release energies (not temperatures) have been observed previously~\cite{cesiumbec-lowreleaseenergy}.  Note that the spin temperatures we report are much lower than those attainable by magnetic trapping or optically pumping a system into a single spin state: even for a fractional population of $10^{-6}$ in other spin states caused by imperfect pumping or spin-flip collisions~\cite{magtrapspintempnote}, the spin temperature in a bias field of 100 mG is 500 nanokelvin, assuming a magnetic moment of one Bohr magneton.  In our experiment, the energy scale is set by the product of the Bohr magneton with the magnetic field \emph{gradient} and the lattice spacing, and it is the relative ease of achieving very small values of this parameter (corresponding to $\mu_B$ times a field of less than one microgauss) which allows us to reach such low spin temperatures.

The mixed-spin region comprises a spectrum of spin excitations, the energy of which can be tuned by adjusting the strength of the gradient~\cite{thermometrypaper}. A spin excitation consists of a spin-up atom swapping places with a spin-down atom on the other side of the zero-temperature spin domain boundary.  Reduction of the gradient reduces the energy of spin excitations.  In the regime of \emph{equilibrated} spins, this causes entropy to flow into the mixed-spin region from other degrees of freedom.  This lowers the temperature of the whole system in a manner locally analogous to standard single-shot adiabatic demagnetization refrigeration~\cite{giauque-ADR,debyeADR}.  The kinetic excitations of a trapped Mott insulator are particle-hole excitations~\cite{gerbier-mott,ho-mott}.  On a microscopic level, particle-hole excitations can couple directly to spin excitations if a particle and a hole on opposite sides of the spin domain boundary annihilate.  Energy may be absorbed by quasiparticles in the first band or by the spin excitations themselves~\cite{doublonpaper}.   In the superfluid phase and during lattice rampup, the kinetic excitations are different and entropy transfer is expected to occur faster and in more complex ways.  Experimentally, it is easier to maintain adiabaticity if most of the path along which the gradient is changed is in a regime of fast relaxation times (e.g. the lower path in Fig.~1a). Thermodynamically, this adiabatic cooling method is a redistribution of entropy from the kinetic degrees of freedom to the entropy which results from partially mixing the two initially segregated spin domains.

It is easily possible for the mixed region to absorb nearly all of the entropy of the system.  In a one-component harmonically trapped Mott insulator which is at a temperature low enough for the particle-hole approximation to hold,  the approximate total entropy is  $k_\mathrm{B} \ln(2)$ times the volume of the ``shells'' between the Mott plateaux~\cite{ho-mott}.  The maximum entropy of the the mixed region is realized when, at low gradient, it is broadened to a substantial fraction of the total size.  In that situation the entropy per site approaches $k_\mathrm{B} \ln(n+1)$, where $n$ is the local number of indistinguishable bosons per site (in our samples, $n$ varies across the trap between 1 and 3). The \emph{maximum} entropy of the spin system is thus larger than the entropy of the kinetic degrees of freedom.  Thus, substantial cooling of the system can be achieved with a single gradient demagnetization ramp. We have made a more quantitative analysis of spin gradient demagnetization by calculating entropy-versus-temperature curves of our system in various field gradients~\cite{demagtheory}.  The results confirm the qualitative argument above and show that spin gradient demagnetization cooling is capable of cooling well below the expected Curie temperature.

Figures~\ref{widthvsgrad}~and~\ref{Tvsgrad} show the results of spin gradient demagnetization cooling experiments.  As the gradient is reduced, the width of the domain wall increases (see Fig.~\ref{widthvsgrad}), indicating transfer of entropy from the kinetic degrees of freedom to the spins.  The width increases much less steeply than would be expected for an isothermal sample, implying cooling.  The observed domain wall width can be converted to a temperature using spin gradient thermometry.  The measured temperature falls rapidly as the gradient is lowered (see Fig.~\ref{Tvsgrad}).  The lowest measured temperature is 350$\pm 50$ picokelvin, making this the coldest implementation of the atomic Mott insulator.  An important goal for future experiments will be to develop an additional method of thermometry to measure the temperature of the kinetic degrees of freedom at this very low temperature scale (for progress towards this goal, see Refs~\cite{greiner-SFMItimescales,blochsinglesite,chinmottplateaux}).  This temperature is colder than the lowest temperature ever measured in an equilibrated kinetic system~\cite{leanhardt-picokelvin}, and it is within a factor of 2 of the expected magnetic ordering temperature~\cite{criticalentropy}.
 
Theoretical curves in Figs.~\ref{widthvsgrad}~and~\ref{Tvsgrad} (see also reference~\cite{demagtheory}) show reasonable agreement with the data.  These curves were fitted to the measured temperatures using only one free parameter: the initial temperature at the maximum gradient.  The initial temperature inferred from this fit is 6.3 nanokelvin.  In our earlier work on thermometry~\cite{thermometrypaper}, where the lowest measured temperature was 1 nanokelvin, some adiabatic demagnetization cooling may have occurred during the preparation of the system.  The flattening-out observed in the measured data at low gradients could be a signal that all \emph{accessible} entropy has been pumped into the spin system.

There are both practical and theoretical limits on the temperatures which can be attained with spin gradient demagnetization cooling.  In traditional magnetic refrigeration experiments, the minimum temperature is often set by the minimum achievable magnetic field or the presence of internal fields in the refrigerant~\cite{nucleardemagreview}.  Analogues of both these limits are relevant here.  Practically, the ratio between the highest and lowest attainable magnetic field gradients is an upper bound on the ratio between the initial and final temperatures.  In our experiment, the maximum value of $\nabla |\textbf{B}_i| / \nabla |\textbf{B}_f|$ is about 1000 (limited mainly by the accuracy of determining the zero crossing of the gradient), which would give a minimum temperature below 10 picokelvin.  Another limit stems from the small difference between the inter-spin interaction energy $U_{\uparrow\downarrow}$ and the mean of the intra-spin interaction energies $(U_{\uparrow\uparrow}+U_{\downarrow\downarrow})/2$.  For $^{87}$Rb, this limit is not expected to preclude cooling below the expected magnetic ordering temperature~\cite{demagtheory}.  The fundamental limit is set by the total entropy of the system.  A single Mott insulator has an entropy per particle much less than $k_B \ln 2$.  At high gradient, the spin entropy is negligible.  Lowering the gradient provides a controlled way of creating a completely mixed two component system at the same entropy which is low enough for spin ordering to occur.  This technique thus provides a specific and realistic method of realizing magnetic phase transitions in optical lattices.

Our scheme differs from single-shot adiabatic demagnetization refrigeration (including that demonstrated in a gas of chromium atoms~\cite{pfaudemag}) in that the magnetic field is replaced by a magnetic field gradient, and spin flip collisions by spin transport.  The chromium scheme cannot be applied to alkali atoms due to the much slower spin-flip rates, and extending it from microkelvins to picokelvins would require sub-microgauss magnetic field control.  In previous work~\cite{thermometrypaper}, we suggested that adiabatic reduction of the gradient could be used for cooling, and some aspects of this proposal have recently been theoretically addressed and verified~\cite{natu-domainwalldynamics}.

The concept behind spin gradient demagnetization cooling is compelling; if adiabaticity can be maintained, then strong cooling in a lattice will occur.  Our experimental implementation was designed to allow the system to equilibrate as much as possible at low lattice depths.  We have tested reversibility by replacing the single gradient ramp by a sequence of ramp down, ramp up, ramp down.  This led to no detectable difference in the final measured temperature.  This indicates that the gradient ramps are adiabatic.  Equilibration in the Mott insulator is more difficult to demonstrate, although the previously demonstrated agreement between spin gradient thermometry and cloud size thermometry at high temperatures~\cite{thermometrypaper} indicates that the kinetic and spin degrees of freedom are equilibrated in that regime.  The fact that the spin distribution fits well to the form expected of an equilibrated spin system is also evidence for equilibration, as is the one-parameter fit to our theoretical predictions (which assume adiabaticity).  However, if the lattice is deepened, then lowered to zero, then raised again, heating is generally observed.  Thus, we cannot rule out the existence of long-lived metastable excitations in the Mott insulating state which do not couple to the spin degrees of freedom and thus do not influence our temperature measurement.  Other experiments have seen evidence of long equilibration times in the Mott insulator~\cite{chicagoMIpaper,doublonpaper}.  For quicker equilibration, spin gradient demagnetization cooling could be implemented with lighter atomic species (e.g. $^7$Li or $^4$He$^*$) and/or shorter period optical lattices.

The cooling technique using isolated spins presented here has achieved spin temperatures and entropies well below the critical values for magnetic ordering, and spin gradient demagnetization cooling of equilibrated spins has cooled to a point within reach of the critical values. This work thus opens a realistic path towards observation of superexchange-driven phase transitions in optical lattices, and extends the potential of ultracold atoms trapped in optical lattices to be used as flexible quantum simulators of strongly interacting many-body systems.

We acknowledge discussions with Eugene Demler, Takuya Kitagawa, David Pekker, and Aditi Mitra.  We thank Aviv Keshet for a critical reading of the manuscript.  H.M. acknowledges support from the NDSEG fellowship program.  This work was supported by the NSF, through a MURI program, and under ARO Grant No. W911NF-07-1-0493 with funds from the DARPA OLE program.  


\begin{thebibliography}{10}
\bibitem{magtrapspintempnote}
A part-per-million spin impurity level would for example be present in a 100 Hz
  magnetic trap with a 100 second lifetime due to dipolar relaxation.
\end{thebibliography}

\end{document}